\documentclass[prl,twocolumn,showpacs,preprintnumbers,amsmath,amssymb,superscriptaddress,amsmath]{revtex4-1}
\makeatletter
\renewcommand*\env@matrix[1][c]{\hskip -\arraycolsep
  \let\@ifnextchar\new@ifnextchar
  \array{*\c@MaxMatrixCols #1}}
\makeatother

\usepackage{amsmath,array}
\usepackage{graphicx,subfigure}  
\usepackage{dcolumn}                 
\usepackage{bm}                          
\usepackage{multirow}

\def\nuebar{\bar{\nu}_{\lowercase{e}}}
\def\nubar{\bar{\nu}}
\graphicspath{{./figures/}}

\begin{document}

  
\title{A proposed search for a fourth neutrino with a PBq anti-neutrino source}

\author{Michel Cribier}
\affiliation{Commissariat \`a l'Energie Atomique et aux Energies Alternatives,\\
Centre de Saclay, IRFU, 91191 Gif-sur-Yvette, France}
\affiliation{Astroparticule et Cosmologie APC, 10 rue Alice Domon et
  L\'eonie Duquet, 75205 Paris cedex 13, France}

\author{Maximilien Fechner}
\affiliation{Commissariat \`a l'Energie Atomique et aux Energies Alternatives,\\
Centre de Saclay, IRFU, 91191 Gif-sur-Yvette, France}

\author{Thierry Lasserre}
\email{Corresponding author: thierry.lasserre@cea.fr}
\affiliation{Commissariat \`a l'Energie Atomique et aux Energies Alternatives,\\
Centre de Saclay, IRFU, 91191 Gif-sur-Yvette, France}
\affiliation{Astroparticule et Cosmologie APC, 10 rue Alice Domon et
  L\'eonie Duquet, 75205 Paris cedex 13, France}

\author{Alain Letourneau}
\affiliation{Commissariat \`a l'Energie Atomique et aux Energies Alternatives,\\
Centre de Saclay, IRFU, 91191 Gif-sur-Yvette, France}

\author{David~Lhuillier}
\affiliation{Commissariat \`a l'Energie Atomique et aux Energies Alternatives,\\
Centre de Saclay, IRFU, 91191 Gif-sur-Yvette, France}

\author{Guillaume Mention}
\affiliation{Commissariat \`a l'Energie Atomique et aux Energies Alternatives,\\
Centre de Saclay, IRFU, 91191 Gif-sur-Yvette, France}

\author{Davide Franco}
\affiliation{Astroparticule et Cosmologie APC, 10 rue Alice Domon et
  L\'eonie Duquet, 75205 Paris cedex 13, France}

\author{Vasily Kornoukhov}
\affiliation{ITEP, ul. Bol. Cheremushkinskaya, 25, 117218 Moscow, Russia}

\author{Stefan Sch\"onert}
\affiliation{Physik Department, Technische Universit\"at M\"unchen, 85747 Garching, Germany}

\date{\today}

\begin{abstract}
Several observed anomalies in neutrino oscillation data can be explained 
by a hypothetical fourth neutrino separated 
from the three standard neutrinos by a squared mass difference 
of a few eV$^2$.  We show that this hypothesis can be tested with 
a PBq (ten kilocurie scale) $^{144}$Ce or $^{106}$Ru antineutrino $\beta$-source deployed at the center of 
a large low background liquid scintillator detector.  
In particular, the compact size of such a source could yield an
energy-dependent oscillating pattern in event spatial distribution
that would unambiguously determine neutrino mass differences and mixing
angles.
\end{abstract}

\maketitle

Most results from neutrino experiments over the last twenty years can
be quite accurately described by a model of oscillations between three
$\nu$ flavors ($\nu_e$, $\nu_{\mu}$, $\nu_{\tau}$) that are
mixtures of three massive neutrinos ($\nu_1$, $\nu_2$, $\nu_3$) separated by squared mass
differences of $\Delta m_{21}^2 = 8 \cdot 10^{-5}~{\rm eV}^2$ and
$\Delta m_{31}^2 = 2.4  \cdot 10^{-3}~{\rm eV}^2$~\cite{pdg}.  
In the past the LSND experiment suggested the existence of a fourth
massive neutrino with a mass of $\sim$1~${\rm eV}^2$~\cite{LSND}.
This evidence has been confirmed by the MiniBoone experiment in the
antineutrino sector~\cite{MiniBoone}.
Recently the hypothetical existence of a fourth $\nu$ has been revived
by a new calculation~\cite{Mueller2011} of the rate of $\nuebar$ production
by nuclear reactors that yields a $\nu$ flux about 3\% higher than
previously predicted. This calculation then implies~\cite{RAA11} that
the measured event rates for all reactor $\nuebar$ experiments
within 100~meters of the reactor are about~6\% too low.  
The deficit can also be explained by a hypothetical fourth massive $\nu$
separated from the three others by $|\Delta m_{\rm
  new}^2|>0.1$~eV$^2$. This mixing can explain  a similar deficit in
the rate of $\nu$ interactions in Gallium solar-$\nu$  detectors when
exposed to artificial $^{51}$Cr and $^{37}$Ar MCi
sources~\cite{GalAno}. 
Combination of~\cite{RAA11} and~\cite{GalAno} deficits is significant at the 99.8\%
C.L., though no conclusive model can explain all data~\cite{GiuntiSterileFit}.
\begin{figure}[!b]
\begin{center}
\includegraphics[scale=0.7]{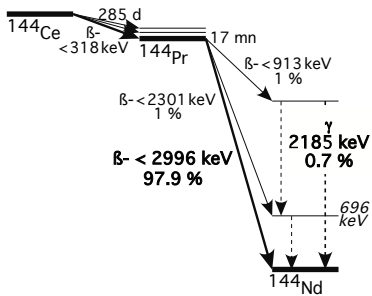}
\caption{\label{f:cepr144} 
Simplified decay scheme of the pair $^{144}$Ce-$^{144}$Pr.
}
\end{center}
\end{figure}
In this letter we propose an unambiguous search for this
fourth neutrino by using a 1.85~PBq (50 kCi) antineutrino source deployed at
the center of a kilo-ton scale detector such as
Borexino~\cite{BorexDet}, KamLAND~\cite{KamLANDSol}, or
SNO+~\cite{SNO+}.
Antineutrino detection will be made via the inverse beta-decay (IBD) reaction
$\bar{\nu}_e$ + p $\rightarrow$ $e^+$+n. The delayed coincidence between
detection of the positron and the neutron capture gamma rays will
allow for a nearly background free experiment. The small size
($\sim$10~g) of the source compared to the size a nuclear reactor
core may allow the observation of the characteristic $\nu$-oscillation
pattern of event positions. \\

Large liquid scintillator (LS) detectors, called~{\rm
  LLSD} hereafter, share key features well suited to search for an
eV-scale fourth $\nu$ ($\nubar$) state mixing with $\nu_e$ ($\nuebar$). The
active mass is composed of about thousand tons of ultra-pure LS
contained in a nylon or acrylic vessel. 
The scintillation light is detected via thousands of photomultipliers
uniformly distributed on a stainless steel spherical vessel.
In Borexino and KamLAND the target is surrounded by mineral oil or scintillator contained in a stainless
steel vessel. This buffer is enclosed in a water tank instrumented by
photomultipliers detecting the Cherenkov light radiated by cosmic
muons. 
In the following we study the deployment of a $\nu$ source of energy
spectrum $\mathcal{S}(E_{\nu})$, mean lifetime  $\tau$, and initial
activity $\mathcal{A}_0$, encapsulated inside a thick tungsten (W) and
copper (Cu) shielding sphere, at the center of a~{\rm LLSD}. We consider a
running time $t_e$ with a fully efficient detector. 
The theoretical expected number of interactions at a radius R and energy $E_{\nu} $ can be written: 
\begin{equation}
\frac{d^2N (R,E_{\nu})}{dR dE_{\nu}} = \mathcal{A} _0  \cdot n
\cdot \sigma(E_{\nu}) \cdot \mathcal{S}(E_{\nu})  \cdot \mathcal{P} (R,E_{\nu})
\int _{0}^{t_e} e^{-t/\tau}dt , 
\label{nevt}
\end{equation}
where $n$ is the density of free protons in the target for inverse beta decay,  
$\sigma$ is the cross section. 
$\mathcal{P} (R,E_{\nu})$ is the 2-$\nu$ oscillation survival probability, defined as:
\begin{equation}
\mathcal{P} (R,E_{\nu}) = 1-\sin^2(2\theta_{new})\cdot\sin^2\left(1.27\frac{\Delta
m_{new}^2[{\rm eV}^2]R[{\rm m}]}{E_{\nu}[{\rm MeV}]}\right), 
\label{prob}
\end{equation}
where $\Delta m_{\rm new}^2$ and~$\theta_{\rm new}$ are the new oscillation
parameters relating $\nu_e$ to the fourth $\nu$. 
In our simulations we assume a 15 cm vertex resolution
and a 5\% energy resolution.
In the no-oscillation scenario we expect a constant $\nu$ rate in
concentric shells of equal thickness (see Eq.~\ref{nevt}).
At 2~MeV the oscillation length is 2.5~m for $\Delta m_{\rm new}^2$= 2
eV$^2$, proportional to $1/\Delta m_{\rm new}^2$ (see Eq.~\ref{prob}). 
A definitive test of the reactor antineutrino anomaly, independent of
the knowledge of the source activity, would be the observation of the
oscillation pattern as a function of the $\nu$ interaction radius
and possibly the $\nu$ energy. \\

Intense man-made $\nu$ sources were used for the calibration of
solar-$\nu$ experiments. In the nineties, $^{51}$Cr ($\sim$750~keV,
$\mathcal{A}_0 \sim$MCi) and $^{37}$Ar (814~keV,
$\mathcal{A}_0$=0.4~MCi) were used as a check of the radiochemical
experiments Gallex and Sage~\cite{SolCalib90s}.
There are two options for deploying $\nu$ sources in LS:
monochromatic $\nu_e$ emitters, like $^{51}$Cr or $^{37}$Ar, or
$\nuebar$ emitters with a continuous $\beta$-spectrum.
In the first case, the signature is provided by $\nu_e$ elastic
scattering off electrons in the LS molecules. This signature
can be mimicked by Compton scattering induced by radioactive and
cosmogenic background, or by Solar-$\nu$  interactions. The
constraints of an experiment with $\nu_e$ impose the use of
a very high activity source (5-10~MCi) outside of the detector
target. In the second option, $\nuebar$ are detected via inverse beta
decay. Its signature, provided by the $e^+$-n delayed
coincidence, offers an efficient rejection of the mentioned
background. For this reason, we focus our studies on $\nuebar$ sources.\\

A suitable $\bar{\nu}_e$ source must have $Q_\beta>$1.8~MeV (the IBD
threshold) and a lifetime that is long enough ($\gtrsim$1~month) to allow for
production and transportation to the detector. For individual nuclei,
these two requirements are contradictory so we expect candidate
sources to involve a long-lived low-$Q$ nucleus that decays to a short-lived
high-$Q$ nucleus. We identified four such pairs
$^{144}$Ce-$^{144}$Pr  ($Q_\beta$(Pr)=2.996~MeV), 
$^{106}$Ru-$^{106}$Rh ($Q_\beta$(Rh)=3.54~MeV), 
$^{90}$Sr-$^{90}$Y ($Q_\beta$(Y)=2.28~MeV), 
and $^{42}$Ar-$^{42}$K  ($Q_\beta$ (K)=3.52~MeV), some of them also
reported in~\cite{Kor94}.
The first three are common fission products from nuclear reactors that
can be extracted from spent fuel rods. While not minimizing the
difficulty of doing this, the nuclear industry does have the
technology to produce sources of the appropriate intensity, at the ppm
purity level. In fact, 10~kCi $^{90}Sr$ sources have been produced and
used industrially for heat generation. Delays obtaining
authorizations for transportation and deployment of the source into
an underground laboratory should be addressed at the start of
the project. 

For this paper, we concentrate on the $^{144}$Ce source
(Fig. \ref{f:cepr144}) because its $Q_\beta$ is greater than that of
$^{90}$Sr and because it is easier to extract chemically than
$^{106}$Ru. We note also that it has a very low production rate of
high-energy $\gamma$ rays ($>1MeV$) from which the $\nuebar$ detector
must be shielded to limit background events.  Finally cerium is present in
fission products of uranium and plutonium at the level of a few percent.
\begin{center}
\begin{table*}[htb!]
\medskip
\begin{tabular}{c|c|c|l|l|l|l|c|c}
  \hline\hline
 Source & F.Y. $^{235}$U/$^{239}$Pu  & $t_{1/2}$ & 1$^{\rm st}$
 $\beta ^-$ (keV)& 2$^{\rm nd}$ $\beta ^-$ (keV) &  $I_{\gamma>1MeV}$ &$ I_{\gamma>2MeV}$&
 W/kCi & kCi/4\,10$^4$ int./y\\
\hline\hline
\hline
 \multirow{3}{*}{$^{144}$Ce-$^{144}$Pr} &
 \multirow{3}{*}{5.2\%/3.7\%} & \multirow{3}{*}{285 d}  & 318 (76\%) &
 &  & &  \multirow{3}{*}{7.47} & \multirow{3}{*}{43.7}\\
 & & & 184 (20\%) &2996 (99\%) & 1380 (0.007\%) &  2185  (0.7\%) & & \\
 & & & 238 (4\%) & 810 (1\%) & 1489 (0.28\%)& & & \\
\hline
\hline
 \multirow{4}{*}{$^{106}$Ru-$^{106}$Rh} &  \multirow{4}{*}{0.5\%/4.3\%} &
  \multirow{4}{*}{373 d}  &
 \multirow{4}{*}{39.4 (100\%)} & 3540 (78\%)
 & 1050 (1.6\%) &  & \multirow{4}{*}{8.40} &\multirow{4}{*}{23.0}\\
 & & & & 3050 (8\%)  &1128-1194 (0.47 \%) & 2112 (0.04\%)  & & \\
 & & & & 2410 (10\%) & 1496-1562 (0.19 \%) & 2366 (0.03\%) & & \\
 & & & & 2000 (2\%)  & 1766-1988 (0.09 \%) & 3400 (0.016\%) & & \\
\hline\hline
\end{tabular}
\caption{\label{tab:sources} Features of $^{144}$Ce-$^{144}$Pr
  and $^{106}$Ru-$^{106}$Rh pairs, extracted from spent nuclear
  fuel. F.Y. are the fission yields of $^{144}$Ce and
  $^{106}$Ru, $t_{1/2}$, their half-lives. $\beta$-end-points
  are given for 1$^{\rm st}$ and 2$^{\rm nd}$ nucleus of each
  pair. The $I_{\gamma}$'s are the branching ratio of gammas $\gamma$
  rays per beta-decay above 1 and 2~MeV.
The two last columns are the heat produced/kCi and the activity required to get 40,000 events/year.
}
\end{table*}
\end{center}
We now focus on the unique oscillation signature induced by
an eV-scale sterile $\nu$ at the center of a LLSD. 
\begin{figure}[!t]
\begin{center}
\includegraphics[scale=0.41]{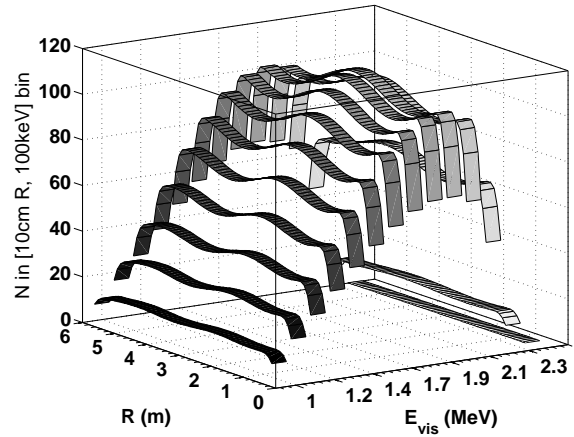}
\caption{\label{f:signal} Advantage of $\nuebar$ sources
  providing both R and E$_{vis}$ oscillation patterns. IBD rate for a 50~kCi
  $^{144}$Ce source deployed at a center of a LLSD, in 10 cm radius bins
  and 100~keV bins of visible energy, E$_{vis}$=E$_e$+2m$_e$. In one
  year, 38,000 $\nuebar$ interact between 1.5~m  and 6~m radius, for
  $\Delta m_{\rm new}^2=2$~eV$^2$ and $\sin^2(2\theta_{\rm
    new})=0.1$. 
}
\end{center}
\end{figure}
For $^{144}$Ce-$^{144}$Pr, 1.85 PBq (50 kCi) source is needed to reach 40,000
interactions in one year in a LLSD, between 1.5 and 6~m away from the
source (n$_H$=5.3 $\cdot$ 10$^{28}$ m$^{-3}$).
This is realized with 14~g of $^{144}$Ce, whereas the total mass of all
cerium isotopes is~$\sim$1.5 kg, for an extraction from selected fission products.
The compactness of the source, $<$4~cm, is small
enough to be considered as a point-like source for $\Delta m_{\rm new}^2$~eV$^2$ oscillation
searches. This source initially releases $\sim$300 W, and it could
be cooled either by convective exchanges with the LS, or via
conduction though an ultrapure copper cold finger connecting the
massive passive shield to a low temperature bath.
$\beta^-$-decay induced $\nuebar$ are detected through IBD. The cross section is
$\sigma(E_e) = 0.956 \,  10^{-43}\times p_e E_e \,\,
\rm{cm}^2$, where $p_e$ and $E_e$ are the momentum
and energy (MeV) of the detected $e^+$, neglecting
recoil, weak magnetism, and radiative corrections (\%-level correction). 
The $e^+$ promptly deposits its kinetic energy in the LS and annihilates emitting two 511~keV $\gamma$-rays,
yielding a prompt event, with a visible energy of E$_e$=
E$_\nu$-($m_n$-$m_p$)~MeV; the emitted keV neutron is captured on a free proton
with a mean time of a few hundred microseconds, followed by the emission of a
2.2~MeV deexcitation $\gamma$-ray providing a delayed coincidence
event. The expected oscillation signal for $\Delta m_{\rm new}^2=2$~eV$^2$
and $\sin^2(2\theta_{\rm new})=0.1$, is shown on Fig.~\ref{f:signal}. 
LLSD are thus well suited to search for an eV-scale fourth $\nu$ state. 
Note that a study of signals of a $^{90}$Sr MCi source external of a
LLSD was done in~\cite{Ian99}.\\

The space-time coincidence signature of IBD
events ensure an almost background-free detection. Backgrounds are
of two types, those induced by the environment or detector, and those
due to the source and its shielding. 
 
The main concern is accidental coincidences between a prompt
(E$>$0.9 MeV) and a delayed energy depositions (E$>$2.0 MeV) occurring within a
time window taken as three neutron capture lifetimes on hydrogen 
(equivalent to about 772 $\mu$sec), and within a volume of 10 m$^3$
(both positions are reconstructed, this last cut leading to a background
rejection of a factor 100).
\begin{figure}[!t]
\begin{center}
\includegraphics[scale=0.41]{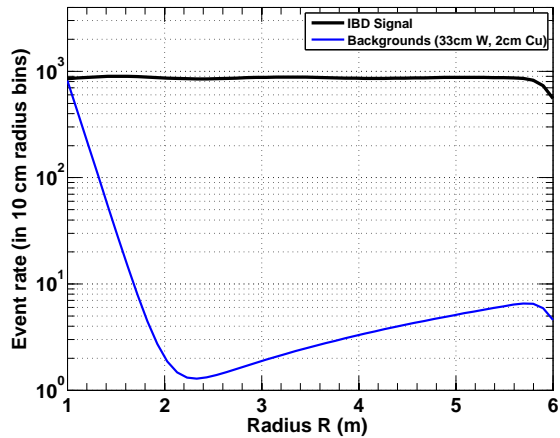} 
\caption{\label{f:signalbkg} 
Signal rate of a 50~kCi $^{144}$Ce deployed for 1 year at a center of a LLSD (black),
compared to the sum of all identified backgrounds rate (blue), as a
function of the detector radius in 10 cm concentric bins.
A shield made of 33 cm of W and 2 cm of Cu attenuates the backgrounds,
dominated primarily by $^{144}$Ce $\gamma$ lines, then by external
bremsstrahlung of the $^{144}$Ce $\beta$-decay electrons slowing down
in the Cerium material.}
\end{center}
\end{figure}
The main source of detector backgrounds originates from
accidental coincidences, fast neutrons, and the long-lived muon induced
isotopes $^9$Li/$^8$He and scales with $R^2$ when using concentric $R$-bins. 
These components have been measured in-situ for the Borexino
geo-$\nu$ searches~\cite{BorexGeo}, at 0.14$\pm$0.02
counts/day/100 tons. Being conservative we increase it to
10 counts/day/100 tons in our simulation.

Geologic $\nuebar$ arising from the decay of radioactive isotopes of
Uranium/Thorium in the Earth have been measured at a rate of a few 
 events/(100 ton.year) in KamLAND~\cite{KamLANDGeo} and Borexino~\cite{BorexGeo}. 
Reactor $\nuebar$ emitted by the $\beta$-decays of the fission
products in the nuclear cores have been measured in KamLAND at a rate of
$\sim$10~events/(100 ton.year) in the energy range of interest~\cite{KamLANDGeo}. 
We use a rate of 20~events/(100 ton.year), which is negligible with
respect to the $\nuebar$ rate from a kCi source. 

The most dangerous source background originates from the energetic
$\gamma$ produced by the decay through excited states of $^{144}$Pr (Table~\ref{tab:sources}).  
We approximate $\gamma$ ray attenuation in a shield of 33~cm of W and
2~cm of Cu with an exponential attenuation
law accounting for Compton scattering and photoelectric effect.
The intensity of 2185~keV $\gamma$ rays is decreased by a factor
$<10^{-12}$ ($\lambda _W \sim$1.2~cm)~\cite{Nist}, to reach a tolerable rate. 

The energy spectrum of external bremsstrahlung photons in the
cerium is estimated with a simulation using the cross
section of~\cite{KM1959}. Results were confirmed with a
GEANT4~\cite{geant4} simulation. The number of photons above a prompt
signal threshold of 0.9 MeV is $6.5 \cdot 10^{-3}$ photons per
$\beta$ decay, and $10^{-4}$ photon per $\beta$ decay $>$2.0~MeV. 

An important remaining background source could be the W shield
itself. Activities at the level of ten to hundreds mBq/kg have been
reported. We anticipate the need of an external layer of ultrapure
copper, set to 2~cm in our simulation. It allows one to achieve the
radiopurity and material compatibility requirements. Assuming a
$\sim$4 tons shield we consider a prompt and delayed event rates of
50~Hz and 25~Hz, respectively. 
The escaping $\gamma$ are attenuated in the LS, assuming
a 20~cm attenuation length~\cite{Nist}. Beyond a distance of 1.5~m from the
source the backgrounds become negligible.
Any of the bremsstrahlung photons or shielding backgrounds can
account for either the prompt or delayed event, depending on their energy. 
The sum of the backgrounds integrated over their energy spectrum is
shown on Fig.~\ref{f:signalbkg}, supporting the case of kCi $\nuebar$
source versus MCi $\nu_e$ source for which solar-$\nu$'s become an
irreducible background.
A light doping of the LS with gadolinium or an oil buffer surrounding the shielding would further
suppress backgrounds; finally, non-source backgrounds could be
measured in-situ during a blank run with an empty shielding.\\

We now assess the sensitivity of an experiment with a 50~kCi $^{144}$Ce
source running for 1~year. 
\begin{figure}[bh!]
\begin{center}
\includegraphics[scale=0.41]{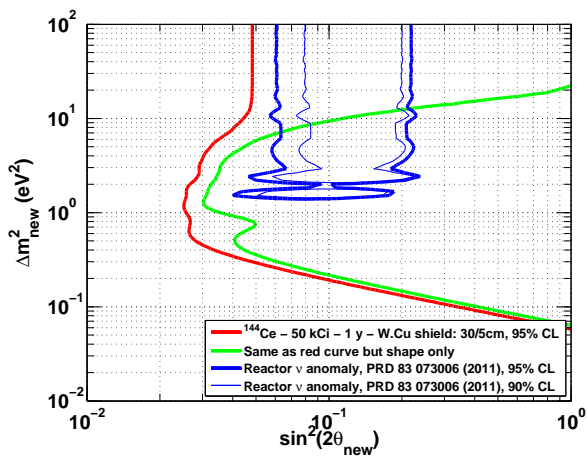}
\caption{\label{f:sensitivity} 95\% C.L. exclusion limit of the 50
  kCi.y $^{144}$Ce experiment obtained in the $\Delta m_{\rm new}^2$
  and $\sin^2(2\theta_{\rm new})$ plane (2~dof). Our result (red and
  green lines with and w/o knowledge of source activity) is
  compared to the 90\% and 95\% C.L. inclusion domains given by the combination
  of reactor neutrino experiments, Gallex and Sage calibration sources
  experiments, MiniBooNE as described in Fig. 8 of~\cite{RAA11}  (blue lines).}
\end{center}
\end{figure} 
With the shield described above and using events between
1.5~m and 6~m, the background is negligible. With $\Delta
m_{\rm new}^2=2$~eV$^2$ and $\sin^2(2\theta _{\rm new})=0.1$, the interaction 
rate decreases from 40,000 to 38,000 per year. 
The 95\% C.L. sensitivity is extracted through the following function:
\begin{equation}
\label{chi2estimator}
\chi^2 = \sum_i \sum_j
\frac{\left( N_{\rm obs}^{i,j}-(1+\alpha)N_{\rm exp}^{i,j}\right)^2}{N_{\rm
    exp}^{i,j}(1+\sigma_b^2 N_{\rm exp}^{i,j})} + \left(
  \frac{\alpha}{\sigma_N}\right)^2,
\end{equation}
where $N_{\rm obs}^{i,j}$ are the simulated data in the no-oscillation case and
 $N_{\rm exp}^{i,j}$ the expectations for a given oscillation scenario, in each energy $E_i$ and radius $R_j$ bin. 
$\sigma_b$~is a 2\% fully uncorrelated systematic error, accounting
for a fiducial volume uncertainty of 1\% in a calibrated
detector, as well as for ($e^+$, n)  space-time coincidence detection
efficiencies uncertainties at the sub-percent level. 
$\sigma_N$ is a normalization error of 1\%,
describing for the source activity uncertainty (from calorimetric
measurement, see~\cite{Kor97}), and $\alpha$ is the associated
nuisance parameter. 
Fig.~\ref{f:sensitivity} clearly shows that 50~kCi of $^{144}$Ce
allows us to probe most of the reactor antineutrino anomaly parameter
space~\cite{RAA11} at 95\% C.L. An analysis assuming no knowledge on
the source activity shows that the oscillatory behavior can be established for $\Delta m_{\rm new}^2 <10$~eV$^2$.
We note that a 10~kCi source would be enough to test the anomaly at
90\%~C.L.\\

%
The reactor $\nuebar$ anomaly and the hypothetical existence of a
4$^{th}$ $\nu$ state mixing with the $\nu_e$ could be
efficiently tested by deploying a $\beta$-source of about 10 grams of
$^{144}$Ce (or $^{106}$Ru) at the center of a large low background
liquid scintillator detector. 
The technical challenge lies in the production of the source
itself, as well as in the realization of a thick ultra pure W/Cu
shielding surrounding the Ce (Ru) material. 
Significant collaborative work would be needed to bring this idea to fruition.
We thank A.~Ianni, M.~Pallavicini, B. Littlejohn,  and J.~Rich for
discussions, and support by the "Origin and Structure of the Universe: Cluster of Excellence for Fundamental Physics"

\end{document}